\documentclass[12pt, a4paper]{article}
\usepackage[cp1251]{inputenc}
\usepackage[english]{babel}
\usepackage{indentfirst}
\usepackage{misccorr}
\usepackage{graphicx}
\usepackage{amsmath, latexsym, amssymb, bm, array, graphics, amsfonts, amsthm, eucal}
\graphicspath{{Sevda1/}}

\makeatletter
\renewcommand{\@biblabel}[1]{#1.\hfill}
\newcommand{\Res}{\mathop{\rm Res\,}}

{

\makeatother

\begin{document}

\newcommand{\mc}[1]{\mathcal{#1}}
\newcommand{\E}{\mc{E}}

\begin{center}
{\bf THE ANALYTICAL SOLUTION OF THE PROBLEM ON PLASMA OSCILLATIONS IN HALF-SPACE
WITH SPECULAR BOUNDARY CONDITIONS}
\end{center}

\centerline{\bf \copyright \;2017  \quad A. V. Latyshev, S. Sh. Suleymanova}

\begin{center}
  {\it
Faculty of Physics and Mathematics,\\
Moscow State Regional University\\
105005, Moscow, Radio str., 10-A\\
e-mail: avlatyshev@mail.ru, sevda-s@yandex.ru\\}
\end{center}

\noindent The boundary problem about behavior (oscillations) of the electronic plasmas
with arbitrary degree of degeneration of electronic gas in half-space with
specular boundary conditions is analytically solved. The kinetic equation
of Vlasov--Boltzmann with integral of collisions of type BGK (Bhatnagar,
Gross, Krook) and Maxwell equation for electric field are applied. Distribution
function for electrons and electric field in plasma in the form of expansion
under eigen solutions of the initial system of equations are received. Coefficients
of these expansions are found by means of the boundary conditions.

\noindent {{\bf Keywords:}
Vlasov---Boltzmann equation,  Maxwell equation,
frequency of collisions, electromagnetic field, modes of Drude, Debaye, and Van Kampen,
dispersion function, boundary value Riemann problem.

\begin{center}
{INTRODUCTION}
\end{center}

The present paper is a continuation of a series papers devoted to the problem of behavior
non-degenerate and degenerate
(Maxwell) plasma in a half-space, which is set on the border of the external
longitudinal electric field (see, for example,
\cite{Lat2001} -- \cite{Lat2007}). In the works \cite{Lat2001} and \cite{Lat2006b}
considered the case of a degenerate plasma.
In the works \cite{Lat2006a} and \cite{Lat2007} considered the case of a
non-degenerate Maxwell plasma.
In this paper we consider the general case of a plasma with an arbitrary degree
of degeneration of the electronic gas.

The concept of ”plasma” appeared in the works of Tonks and Langmuir for the first
time \cite{Tonks}. The problem of electron
plasma oscillations was considered by A.A. Vlasov (see, for example, \cite{Vlasov}).
The problem of plasma oscillation
turns out to be formulated correctly as a boundary problem of mathematical physics
in the work \cite{Landau46}.

In this paper, the kinetic equation of Vlasov--Boltzmann with integral of collisions
of type BGK (Bhatnagar,
Gross, Krook) and Maxwell equation for electric field are applied for solution the
problem of the behavior of the plasma
with an arbitrary degree of degeneration of the electronic gas in a half-space in
an external alternating longitudinal electric field.

Principal step is the reduction of the boundary problem for one-dimensional and one-velocity.
We use the method of successive approximations, linearization of equations with
respect to the absolute Fermi-Dirac distribution
of electrons and the law of conservation of the number of particles. Then, separation
of variables reduces equations of the problem to the characteristic system of the equations.
In the space of generalized functions are their eigen solutions of the original
system corresponding the continuous spectrum (mode of Van Kampen).

By solving the dispersion equation, we find their eigen solutions, which are
responsible accession and the discrete spectrum (modes of Drude and Debye).
Then the common decision of a boundary problem in the form of expansion of eigen
solutions is formed. The expansion coefficients are determined from the boundary conditions.
This allows for expansion of the distribution function and the electric field explicitly.

It determines the structure of the screened electric field. It turned out that
there is a domain of values parameters of the problem, in which there is no mode of Debye.

It is shown firstly, mode of Drude describing the volume conductivity, exists at
all values of the parameters of the problem.
Secondly, mode of Debye describes screening the electric field and exists for
frequencies of oscillation in the external field which less a some critical
frequency (it is located near the plasma resonance).
And mode Van Kampen \cite{VanKampen} representing a mix of (chaotization)
eigen solutions of the Vlasov--Boltzmann equation also exist for all values of
the parameters of the problem.

\begin{center}
{1. PROBLEM STATEMENT AND THE BASIC EQUATIONS}
\end{center}

For analytical solution such difficult problem as the problem of plasma oscillations,
it must first be reduced to one-dimensional and one-velocity \cite{Lat}.
To do this, the vector of the external electric field is directed along the same axis,
which is orthogonal to the surface layer with the plasma.

Further we pass to the dimensionless variables and parameters. If dimensionless
equations naturally arises small parameter such as a perturbation of the dimensionless
(chemical) potential caused by the presence of an external electric field.
Using the method of small parameter, we linearize the problem relative to the absolute
Fermi--Dirac distribution of the electron.
Applying the law of conservation of the number of particles, we finally formulate
the problem in the form of one-dimensional and one-velocity boundary value problem

$$
\mu\dfrac{\partial H}{\partial x_1}+w_0H(x,\mu)=\mu e(x_1)+
\int_{-\infty}^{\infty}k(\mu',\alpha)H(x,\mu')d\mu',
\eqno{(1.1)}
$$
$$
\dfrac{d e(x_1)}{d x_1}=
\varkappa^2(\alpha)\int_{-\infty}^{\infty}k(\mu',\alpha)H(x_1,\mu')d\mu'.
\eqno{(1.2)}
$$

In the equations (1.1) and (1.2)
$$
\varkappa^2(\alpha)=\dfrac{32\pi^2e^2p_T^3s_0(\alpha)}{(2\pi\hbar)^3m\nu^2},\qquad
w_0=1-i\dfrac{\omega}{\nu}=1-i \omega \tau=1-i\dfrac{\Omega}{\varepsilon},
$$
where $\Omega=\omega/\omega_p$,
$\varepsilon=\nu/\omega_p$, $\omega_p$ is plasma (Langmuir) frequency,
$
\omega_p=\sqrt{4\pi e^2 N/m}.
$
Here $N$ is numerical density (concentration) of electrons in the equilibrium state.

Let's express parameter $\varkappa$  through numerical density.
From definition of a numerical density it follows that
$$
N=\int f_0(P,\alpha)d\Omega_F=\frac{2p^3_T}{(2\pi\hbar)^3}\int\frac{d^3P}{1+e^{P^2-\alpha}}=
\frac{8\pi p^3_T}{(2\pi\hbar)^3}\int_{0}^{\infty}
\frac{P^2dP}{1+e^{P^2-\alpha}}=\frac{8\pi p^3_T}{(2\pi\hbar)^3}s_2(\alpha),
$$
where
$$
s_2(\alpha)=\int_{0}^{\infty}\frac{P^2dP}{1+e^{P^2-\alpha}}=\frac{1}{2}l_0(\alpha),
\quad l_0(\alpha)=\int_{0}^{\infty}\ln(1+e^{\alpha-P^2})dP.
$$

Hence, the numerical density of particles of plasma and thermal wave number
$k_T={mv_T}/{\hbar}$ are related
$$
N=\frac{l_0(\alpha)}{2\pi^2}k^3_T=\frac{s_2(\alpha)}{\pi^2}k^3_T.
$$

Not difficult obtain that
$$
\varkappa^2(\alpha)=\frac{\omega^2_p}{\nu^2} \cdot \frac{s_0(\alpha)}{s_2(\alpha)}=
\frac{\omega^2_p}{\nu^2} \cdot \frac{1}{r(\alpha)}=\dfrac{1}{\varepsilon^2r(\alpha)},
$$
where
$$
r(\alpha)=\dfrac{s_2(\alpha)}{s_0(\alpha)}, \qquad \varepsilon=\dfrac{\nu}{\omega_p}.
$$

It is known that the frequency of the plasma oscillations are usually much more
than the frequency of
collisions between electrons in the metal \cite{Landau10}.Therefore, in the case
when $\omega\sim\omega_p$
the condition $\omega_p \gg \nu$ is performed.

Consider the condition of specular reflection of electrons from the boundary of half-space
$$
f(x=0, v_x, v_y, v_z, t)=f_{eq}(x=0, -v_x, v_y, v_z, t), \qquad v_x>0.
$$

We will linearize this boundary condition, we obtain
$$
H(0, \mu)=H(0, -\mu), \qquad 0<\mu<1.
\eqno{(1.3)}
$$

The boundary condition for the field on the surface of the plasma has the form
$$
e(0)=1,
\eqno {(1.4)}
$$
and away from the surface of the field intended to be limited
$$
e(+\infty) = e_{\infty}, \quad  |e_\infty|<+\infty.
\eqno {(1.5)}
$$

We need the condition of non-permeability of electrons through the plasma boundary
as a boundary condition
$$
\int v_x f(x, \mathbf v, t)d\Omega_F=0.
$$

Hence we obtain the following integral condition
$$
\int_{-\infty}^{\infty}\mu'H(0,\mu')f_0(\mu',\alpha) d\mu' = 0.
\eqno {(1.6)}
$$

Condition (1.6) is the condition of non-permeability of electrons through the plasma
boundary.

\begin{center}
{2. EIGENFUNCTIONS OF THE CONTINUOUS SPECTRUM}
\end{center}

First, we seek the general solution of the system of equations (1.1) and (1.2).

Application of the general Fourier method of the separation of variables in several
steps results in the following substitution
$$
H_\eta(x,\mu)=\mathrm{exp}\left(-\frac{w_0x}{\eta}\right)\Phi(\eta,\mu), \qquad
e_\eta(x)=\mathrm{exp}\left(-\frac{w_0x}{\eta}\right)E(\eta),
\eqno {(2.1)}
$$
where $\eta$ is the spectrum parameter or the parameter of separation which is complex
in general.

We substitute the equalities (2.1) into the equations (1.1) and (1.2). We obtain the
following characteristic
system of equations

$$
(\eta-\mu)\Phi(\eta,\mu)=\eta\mu\frac{E(\eta)}{w_0}+\frac{\eta}{w_0}
\int_{-\infty}^{\infty}k(\mu',\alpha)\Phi(\eta,\mu')d\mu',
\eqno {(2.2)}
$$

$$
-\frac{w_0}{\eta}E(\eta)=\frac{1}{\varepsilon^2r(\alpha)}
\int_{-\infty}^{\infty}k(\mu', \alpha)\Phi(\eta,\mu')d\mu'.
\eqno{(2.3)}
$$

Let us introduce the designations
$$
n(\eta)=\int_{-\infty}^{\infty}k(\mu',\alpha)\Phi(\eta,\mu')d\mu'.
$$

By means of this equality we will rewrite the equations (2.2) and (2.3)
in the form
$$
(\eta-\mu)\Phi(\eta,\mu)=\frac{E(\eta)}{w_0}\mu\eta+\frac{\eta n(\eta)}{w_0},
\eqno {(2.4)}
$$
$$
-\frac{w_0}{\eta}E(\eta)=\frac{n(\eta)}{\varepsilon^2r(\alpha)}.
\eqno {(2.5)}
$$

Let us introduce the designations
$$
\eta^2_1\equiv\eta_1(\alpha)=w_0\varepsilon^2\frac{s_2(\alpha)}{s_0(\alpha)}=
w_0\varepsilon^2r(\alpha)=\varepsilon(\varepsilon-i\Omega)r(\alpha).
$$

From equations (2.4) and (2.5) we obtain the following equation
$$
(\eta-\mu)\Phi(\eta,\mu)=\frac{E(\eta)}{w_0}(\eta\mu-\eta^2_1).
\eqno {(2.6)}
$$

For $\eta\in(-\infty,+\infty)$ we look for a solution of equation (2.6) in the space
of generalized
function \cite{Zharinov}
$$
\Phi(\eta,\mu)=\frac{E(\eta)}{w_0}(\mu\eta-\eta^2_1)P\frac{1}{\eta-\mu}+
g(\eta)\delta(\eta-\mu).
\eqno {(2.7)}
$$

In the equation (2.7) $\eta\in(-\infty,+\infty)$, $\mu\in(-\infty,+\infty)$. The set
of values
$\eta$, filling the real line $-\infty<\eta<+\infty$ is called a continuous spectrum
of the characteristic equation.

In the equation (2.7) $\delta(x)$ is the Dirac delta function, the symbol $Px^{-1}$
denotes the distribution, i.e.
the principal value of the integral of $x^{-1}$, function $g(\eta)$ acts as an arbitrary
"constant"\, of integration.

The solution (2.7) of the equation (2.6) are called the eigenfunctions of the
characteristic equation.

To find the function $g(\eta)$, we substitute (2.7)  in the definition of normalization
functions $n(\eta)$. The result is that
$
g(\eta)={\eta_1^2E(\eta)\Lambda(\eta)}/[{\eta k(\eta,\alpha)}].
$
Here the dispersion function is entered
$$
\Lambda(z)=\Lambda(z,\Omega,\varepsilon,\alpha)=1+\frac{z}{w_0\eta_1^2}
\int_{-\infty}^{\infty}\frac{\eta^2_1-\mu' z}{\mu'-z}k(\mu',\alpha)d\mu'.
\eqno {(2.8)}
$$

Eigenfunctions (2.7) of the characteristic equation (2.6) using (2.8) can be represented as
$$
\Phi(\eta,\mu)=\frac{E(\eta)}{w_0}
\left[P\frac{\mu\eta-\eta^2_1}{\eta-\mu}-w_0\eta_1^2\frac{\Lambda(\eta)}
{\eta k(\eta,\alpha)}\delta(\eta-\mu)\right].
\eqno {(2.9)}
$$

The collection of eigenfunctions $\Phi(\eta,\mu)$ of the characteristic equation
corresponds to the continuous spectrum.
They are often called "mode Van Kampen" (see \cite{VanKampen} and
\cite{Kadomtsev}).

Eigenfunction can be represented as
$$
\Phi(\eta,\mu)=\frac{E(\eta)}{w_0}F(\eta,\mu),
$$
where
$$
F(\eta,\mu)=P\frac{\mu\eta-\eta^2_1}{\eta-\mu}-w_0\eta_1^2
\frac{\Lambda(\eta)}{\eta k(\eta,\alpha)}\delta(\eta-\mu).
$$

The dispersion function problems $\Lambda(z)$ can be represented in the following form
$$
\Lambda(z)=1-\frac{1}{w_0}-\frac{z^2-\eta^2_1}{w_0\eta^2_1}\lambda_0(z,\alpha).
$$
Here function is entered
$$
\lambda_0(z,\alpha)=1+z\int_{-\infty}^{\infty}\frac{k(\mu,\alpha)d\mu}{\mu-z}.
$$
For its boundary values above and below are carried out on the real axis of the
Sokhotzky formulas \cite{Gahov}
$\lambda^\pm_0(\mu,\alpha)=\lambda_0(\mu,\alpha)\pm
i\pi\mu k(\mu,\alpha). $

Using these formulas, we can easily calculate the boundary
values of above and below on the real axis dispersion function of problems
$$
\Lambda^\pm(\mu)=\Lambda(\mu)\pm i\frac{\pi}{w_0\eta^2_1}\mu k(\mu,\alpha)(\eta^2_1-\mu^2),
\quad\frac{\Lambda^+(\mu)+\Lambda^-(\mu)}{2}=\Lambda(\mu).
$$

Fig. 1 and 2 are graphs respectively the real and imaginary parts of the dispersion
function $\Lambda^{+}(\mu)$
in the case $\Omega=1$ and $\varepsilon=0.01$,
curves 1, 2, 3 correspond to the values of the dimensionless chemical potential
$\alpha= 3, 0, -1$.

\begin{center}
{3. ZEROS DISPERSION FUNCTION}
\end{center}

We find the zeros of the dispersion equation
$$
\frac{\Lambda(z)}{z}=0.
\eqno {(3.1)}
$$

It is easy to see that the value of the dispersion function at infinity is equal to
$$
\Lambda_{\infty}=\Lambda(\infty)=1-\frac{1}{w_0}+\frac{1}{w^2_0\varepsilon^2}.
$$
Hence we find that
$$
\Lambda_{\infty}=\frac{-i\nu\omega+\omega^2_p-\omega^2}{(\nu-i\omega)^2}\ne0
$$
for any $\nu\ne0$, i.e. in any collisional plasma.

This means that the point $z_i=\infty$ is a zero dispersion equation. This
point is point of the spectrum associated to the continuous spectrum.
Point $z_i=\infty$ corresponds to the following solution of the original system
of equations (2.5) and (2.6)
$$
H_{\infty}(x,\mu)=\frac{E_{\infty}}{w_0} \cdot \mu, \quad  e_{\infty}=E_{\infty}.
\eqno {(3.2)}
$$

Here $E_{\infty}$ is an arbitrary constant.

The solution (3.2) does not depend on the chemical potential. This solution is
naturally called as mode of Drude. It describes the
volume conductivity of metal, considered by Drude (see, for example, \cite{Ashkroft}).

By definition, the discrete spectrum of the characteristic equation (2.6) is the set
of finite complex zeros
of the dispersion equation (3.1) which do not lie on the real axis (a cut dispersion
function).

\begin{figure}[t]
\begin{center}
\includegraphics[width=0.47\textwidth]{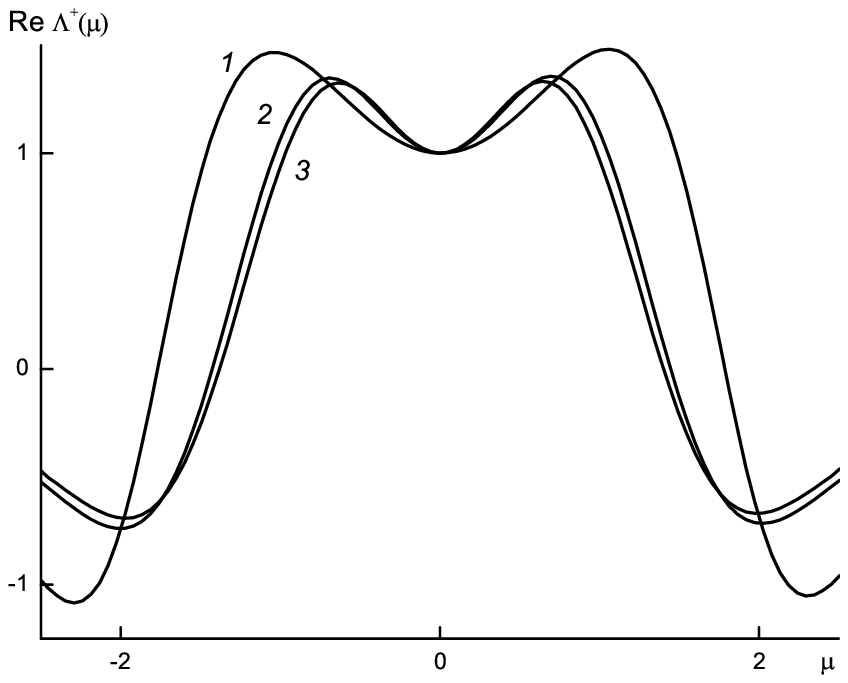}\hfill
\includegraphics[width=0.47\textwidth]{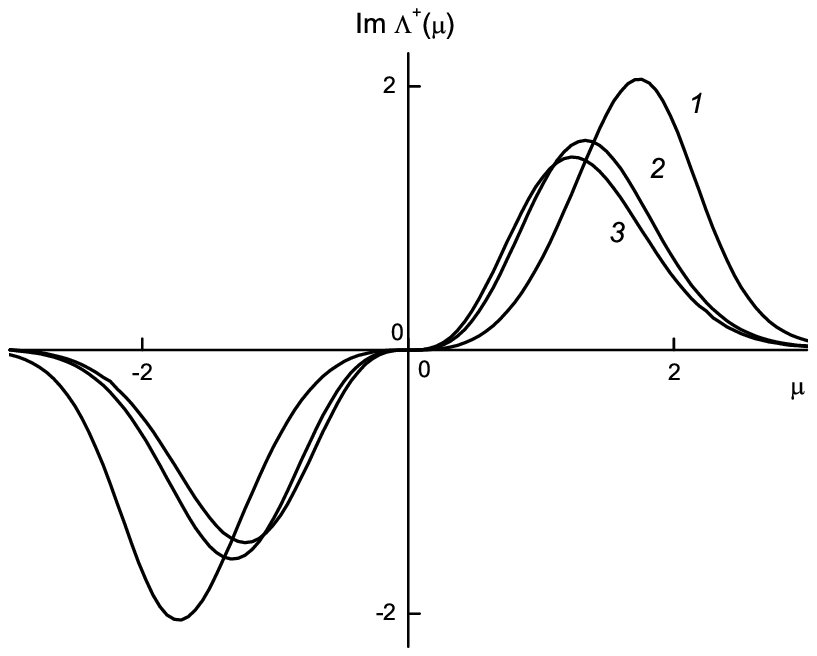}\\
\parbox[t]{0.47\textwidth}{\hspace{3cm}{Fig. 1.}}\hfill
\parbox[t]{0.47\textwidth}{\hspace{3cm}{Fig. 2.}}
\end{center}
\end{figure}

We start to search zeros of the dispersion function.
Let us take Laurent series of the dispersion function
$$
\Lambda(z)=\Lambda_{\infty}+\frac{\Lambda_2}{z^2}+
\frac{\Lambda_4}{z^4}+ \cdot \cdot \cdot , \ \quad z\to\infty.
\eqno {(3.3)}
$$
where
$$
\Lambda_2=\frac{s_4(\alpha)-\eta^2_1s_2(\alpha)}{w_0\eta^2_1s_0(\alpha)}, \ \qquad
\Lambda_4=\frac{s_6(\alpha)-\eta^2_1s_4(\alpha)}{w_0\eta^2_1s_0(\alpha)},
\cdots, \ \quad
s_n(\alpha)=\int_{0}^{\infty}\mu^n f_0(\mu, \alpha)d\mu.
$$

From the expansion (3.3) we see that in a neighborhood of infinity, there are
two zeros $\pm\eta_0$
dispersion function $\Lambda(z)$
$$
\pm\eta_0\approx\sqrt{-\frac{\Lambda_2}{\Lambda_{\infty}}}.
\eqno {(3.4)}
$$
Since the dispersion function is even then its zeros differ from each other by sign.
By zero $\eta_0$ we understand such radical value from (3.4) that
$
\mathrm{Re}({w_0}/{\eta_0})>0.
$
For such a zero exponent $\mathrm{exp}[-({w_0}/{\eta_0})x]$ is monotonically
decreasing as $x\to+\infty$.

Zero $\eta_0$ corresponds to the following equation

$$
H_{\eta_{0}}(x,\mu)=\mathrm{exp}\left(-\frac{w_0}{\eta_0}x\right)\Phi(\eta_0,\mu),
\quad
e_{\eta_0}(x)=\mathrm{exp}\left(-\frac{w_0}{\eta_0}x\right)E_0.
$$
Here
$$
\Phi(\eta_0,\mu)=\frac{E_0}{w_0}\frac{\eta_0\mu-\eta^2_1}{\eta_0-\mu}.
$$

This solution is naturally called the mode of Debye (this is plasma mode).
In the case of low frequencies it describes
well-known screening of Debay \cite{Ashkroft}.

Equality (3.4) we will present in the explicit form
$$
\eta_0=\eta_0(\alpha,\Omega,\varepsilon)\approx\sqrt{\frac{(\Omega+i\varepsilon)^2
[\eta^2_1s_2(\alpha)-s_4(\alpha)]}{w_0\eta^2_1s_0(\alpha)(\Omega^2-1+i\varepsilon \Omega)}}.
\eqno {(3.5)}
$$

From (3.5) we see that near the plasma resonance (when $\Omega\approx1$, i.e.
$\omega\approx\omega_p$) module of zero $|\eta_0(\alpha,\Omega,\varepsilon)|$
becomes unlimited for all values of the dimensionless chemical potential $\alpha$
in the case $\varepsilon\to 0$.

\begin{center}
{4. ON THE EXISTENCE OF PLASMA MODES}
\end{center}

Zero $\eta_0$ is a function of the parameters of the initial system of equations
$\mu$, $\omega$ and $\nu$ or the function of the parameters
$(\alpha,\Omega,\varepsilon)$.
Required to find the domain $D^+(\alpha)$, which lies in the plane of the parameters
$(\Omega,\varepsilon)$, such that if
$(\Omega,\varepsilon)\in D^+(\alpha)$, then the number of zeros $N$ of the dispersion
function
$\Lambda(z)$ is two $N=2$. The $D^-(\alpha)$ denotes a domain in the plane of the
parameters that the number of zeros
of the dispersion function is zero $N=0$. The curve, which is the boundary of
these domains, denoted by $L=L(\alpha)$.

The set of physically significant parameters $(\Omega,\varepsilon)$ fills a quarter-plane
$\mathbb R^2_+ = \{(\Omega,\varepsilon):\Omega\geqslant0,
\varepsilon\geqslant0\}$. Case $\Omega\geqslant0$ (or $\omega=0$) corresponds to the
external stationary electric field,
and case $\varepsilon=0$ (or $\nu=0$) corresponds to the case is responsible
collision-less plasma.

We take the contour $\mathrm\Gamma_{\rho}=C^+_{\rho}\cup C^-_{\rho}$. This contour
consists of two closed semi-circles
$C^+_\rho$ and $C^-_\rho$ of radius $R=1/\rho$
lying in the upper and lower half-planes; $\rho$ is  sufficiently small positive real
number, $C^{\pm}_\rho=\{z=x+iy, |z|=1/\rho,
|x\pm i\rho|\leqslant 1/\rho\}$.
The number $R$ we take large enough to zero of dispersion function
(if they exist) lying inside the domain $D_\rho$ bounded by the contour
$\mathrm\Gamma_\rho$. Note that if  $\rho\to0$ domain  $D_\rho$
passes to $D_0$ bounded by the contour
$\mathrm\Gamma_0=\lim\limits_{\rho\to0}\mathrm\Gamma_\rho$.
This domain coincides with the complex plane with a cut along the real axis.

Then according to the principle of argument the number \cite{Gahov, Rasulov} of zeros
$N$ of the dispersion function
in the domain $D_\rho$ equals to

$$
N=\frac{1}{2\pi i}\oint\limits_{\mathrm\Gamma_\rho}d\ln\Lambda(z).
$$

Considering the limit in this equality when $\rho\to0$ and taking into account
that the dispersion function
is analytic in the neighborhood of the infinity, we obtain that

$$
N=\frac{1}{2\pi i}\int_{-\infty}^{\infty}d\ln\Lambda^+(\tau)-
\frac{1}{2\pi i}\int_{-\infty}^{\infty}d\ln\Lambda^-(\tau).
$$

After some transformations, we obtain that
$$
N=\frac{1}{\pi i}\int_{0}^{\infty}d\ln\frac{\Lambda^+(\tau)}
{\Lambda^-(\tau)}=2\varkappa_{\mathbb R_+}(G).
\eqno {(4.1)}
$$
Here $\varkappa_{\mathbb R_+}(G)$ is the index of the function
$
G(\tau)={\Lambda^+(\tau)}/{\Lambda^-(\tau)},
$
calculated along the positive real axis.

Thus, equality (4.1) means that the number of zeros of the dispersion function
$\Lambda(z)$ is doubled the index function $G(\tau)$
calculated along the positive real axis.

Consider a curve in the complex plane $\mathrm\Gamma_\alpha=\mathrm\Gamma(\alpha)$,
$$
\mathrm\Gamma(\alpha):z=G(\tau), \ 0\leqslant\tau\leqslant+\infty,
$$
It is obvious that $G(0)=1$, $\lim\limits_{\tau\to+\infty}G(\tau)=1$. Consequently,
according to (4.1), the number of values $N$
equals to doubled number of turns of the curve
$\mathrm\Gamma(\alpha)$ around the point of origin, i.e.
$$
N=2\varkappa(G), \quad \varkappa(G)=\mathrm{ind}_{[0,+\infty]}G(\tau).
$$

Let us single real and imaginary parts of the function $G(\mu)$ out. At first,
we represent the function
$G(\mu)$ in the form
$$
G(\mu)=\frac{\Omega^+(\mu)}{\Omega^-(\mu)},
$$
where
$$
\Omega^\pm(\mu)=(w_0-1)\eta^2_1+(\eta^2_1-\mu^2)\lambda_0(\mu,\alpha)\pm
is(\mu,\alpha)(\eta^2_1-\mu^2), \ \ \ \
s(\mu,\alpha)=\frac{\pi}{2s_0(\alpha)}\mu f_0(\mu,\alpha).
$$

Taking into account that
$$
w_0-1=-i\frac{\Omega}{\varepsilon}, \quad
\eta^2_1=\varepsilon r(\alpha)(\varepsilon-i\Omega),\quad
(w_0-1)\eta^2_1=-\Omega(\Omega+i\varepsilon)r(\alpha).
$$

Single out the the real and imaginary parts of the functions $\Omega^\pm(\mu)$. We have
$$
\Omega^\pm(\mu)=-P^\pm(\mu)-iQ^\pm(\mu),
$$
where
$$
P^\pm(\mu)=(1+\gamma)^2r(\alpha)+\lambda_0(\mu,\alpha)
(\mu^2-\varepsilon^2r(\alpha))\mp\varepsilon(1+\gamma)r(\alpha)s(\mu,\alpha),
$$
$$
Q^\pm(\mu)=\varepsilon(1+\gamma)r(\alpha)(1+\lambda_0(\mu,\alpha))\pm
(\mu^2-\varepsilon^2r(\alpha))s(\mu,\alpha).
$$

Now the coefficient $G(\mu)$ can be represented as
$$
G(\mu)=\frac{P^+(\mu)+iQ^+(\mu)}{P^-(\mu)+iQ^-(\mu)}.
$$

We can easily single real and imaginary parts of the function $G(\mu)$ out
$$
G(\mu,\alpha)=\frac{P^+P^-+Q^+Q^-}{(P^-)^2+(Q^-)^2}+i\frac{P^-Q^+-P^+Q^-}{(P^-)^2+(Q^-)^2},
$$
or briefly
$$
G(\mu)=G_1(\mu)+iG_2(\mu),
$$
where
$$
G_1(\mu)=\frac{g_1(\mu)}{g(\mu)}, \quad
G_2(\mu)=\frac{g_2(\mu)}{g(\mu)}.
$$

Here
$$
g(\mu)=[P^-(\mu)]^2+[Q^-(\mu)]^2=
[\Omega^2r(\alpha)+\lambda_0(\mu,\alpha)(\mu^2-\varepsilon^2r(\alpha))+
\varepsilon\Omega r(\alpha)s(\mu,\alpha)]^2+
$$
$$
+[\varepsilon\Omega(1+\lambda_0(\mu,\alpha))-s(\mu,\alpha)
(\mu^2-\varepsilon^2r(\alpha))]^2,
$$ \medskip
$$
g_1(\mu)=P^+(\mu)P^-(\mu)+Q^+(\mu)Q^-(\mu)=
[\Omega^2r(\alpha)+\lambda_0(\mu,\alpha)(\mu^2-\varepsilon^2r(\alpha))]^2+
$$
$$
\hspace{3cm}+\varepsilon^2\Omega^2r^2(\alpha)[(1+\lambda_0(\mu,\alpha))^2-s^2(\mu,\alpha)]-
(\mu^2-\varepsilon^2r(\alpha))^2s^2(\mu,\alpha),
$$ \medskip
$$
g_2(\mu)=P^-(\mu)Q^+(\mu)-P^+(\mu)Q^-(\mu)=
2s(\mu,\alpha)\big\{[\Omega2r(\alpha)+\hspace{3cm}
$$
$$
\hspace{2cm}+\lambda_0(\mu,\alpha)
(\mu^2-\varepsilon^2r(\alpha))](\mu^2-\varepsilon^2r(\alpha))+
\varepsilon^2\Omega^2r^2(\alpha)(1+\lambda_0(\mu,\alpha))\big\}.
$$

We consider (see fig. 3,4) the curve $L_{\alpha}=L(\alpha,\Omega,\varepsilon)$
which is defined in implicit form
by the following parametric equations
$$
L_\alpha=L_\alpha(\Omega,\varepsilon): \quad  g_1(\mu,\alpha,\Omega,\varepsilon)=0, \quad
g_2(\mu,\alpha,\Omega,\varepsilon)=0, \quad 0\leqslant\mu\leqslant+\infty,
$$
and which lays in the plane of the parameters of the problem $(\Omega,\varepsilon)$,
and when passing through this curve the index of the function $G(\mu)$
at the positive semi-axis changes stepwise.

Each curve $L_\alpha$  separates the plane of the parameters $(\Omega,\varepsilon)$
into two domains
$D^+(\alpha)$ and $D^-(\alpha)$, such that if the point $(\Omega,\varepsilon)$ passes
from one domain to another
the index of the function  $G(\mu)$ at the positive semi-axis changes stepwise.

Fig. 3 and 4 is a graph of the curve $L$ which separates the domain $D^{+}$ and $D^{-}$
with the corresponding values of the dimensionless chemical potential $\alpha=-3,3$.

As in the work \cite{Lat1998TMF} we can prove that if
$(\Omega,\varepsilon)\in D^+(\alpha)$
then $\varkappa_{[0,+\infty]}(G)=1$ (the curve L encircles
the point of origin once), and if $(\Omega,\varepsilon)\in D^-(\alpha)$
then $\varkappa_{[0,+\infty]}(G)=0$ (the curve L does not encircle the point of origin).

From the expression (4.1) one can see that the number of zeros of the dispersion
function is two
($N=2$), if $(\Omega,\varepsilon)\in D^+(\alpha)$ and the dispersion function does
not have zeros, if $(\Omega,\varepsilon)\in D^-(\alpha)$.

We note, that in the work \cite{Lat1998TMF} the method of analysis of boundary
regime when $(\Omega,\varepsilon)\in L_\alpha$ was developed.

We deduce explicit parametric equations of the curve $L_\alpha$ which separated
quarter-plane of the parameters $(\Omega,\varepsilon)$
in the two domains $D^+(\alpha)$ and $D^-(\alpha)$.

From the equation $g_2(\mu,\alpha,\Omega,\varepsilon)=0$ we find
$$
\Omega^2=-\frac{1}{r(\alpha)} \cdot
\frac{(\mu^2-\varepsilon^2r(\alpha))\lambda_0(\mu,\alpha)}{\mu^2+
\varepsilon^2r(\alpha)\lambda_0(\mu,\alpha)}.
\eqno {(4.2)}
$$

Consider the equation $g_1(\mu,\alpha,\Omega,\varepsilon)=0$. Let us transform
this equation
with the help of  the equation (4.2). We perform this calculation in general.
From the equation
$
g_1=P^-Q^+-P^+Q^-=0
$
we find $P^-=P^+({Q^-}/{Q^+})$. Further, we find that
$$
 g_1\Big|_{g_2=0}=[P^-Q^+-P^+Q^-]\Bigg|_{P^-=\frac{Q^-}{Q^+}P^+}=
\frac{Q^-}{Q^+}\left[(P^+)^2+(P^-)^2\right].
$$

It is obvious that the equation $g_1\big|_{g_2=0}=0$ is equivalent to the equation
$Q^-(\mu)=0$. From this equation and (4.2) we find the parametric equations of
curves $L_\alpha$:
$$
L_\alpha: \quad \Omega=\sqrt{L_1(\mu)}, \quad
\varepsilon=\sqrt{L_2(\mu)}, \quad 0\leqslant\mu\leqslant+\infty.
\eqno {(4.3)}
$$

In (4.3) we introduce the designations
$$
L_1(\mu)=\frac{s_0(\alpha)}{s_2(\alpha)} \cdot
\frac{\mu^2[\lambda_0(\mu,\alpha)(1+\lambda_0(\mu,\alpha))+s^2(\mu,\alpha)]^2}
{[-\lambda_0(\mu,\alpha)][(1+\lambda_0(\mu,\alpha))^2+s^2(\mu,\alpha)]}
$$
and
$$
L_2(\mu)=\frac{s_0(\alpha)}{s_2(\alpha)} \cdot \frac{\mu^2s_2(\mu,\alpha)}
{[-\lambda_0(\mu,\alpha)][(1+\lambda_0(\mu,\alpha))^2+s^2(\mu,\alpha)]}.
$$

Thus, we have constructed a curve $L_\alpha$ which is the boundary of domains
$D^+(\alpha)$ and
$D^-(\alpha)$.
Recall that if $(\Omega,\varepsilon)\in D^+(\alpha)$ then
$$
\varkappa(G)=\mathrm{ind}_{[0,+\infty]}\frac{\Lambda^+(\mu)}{\Lambda^-(\mu)}=1.
$$
This means that the curve $\mathrm\Gamma_\alpha$ encircles the point of origin once.
And if $(\Omega,\varepsilon)\in D^-(\alpha)$ then
$$
\varkappa(G)=\mathrm{ind}_{[0,+\infty]}\frac{\Lambda^+(\mu)}{\Lambda^-(\mu)}=0.
$$

This means that the curve $\mathrm\Gamma_\alpha$ does not encircle the point of origin.
The curve $\mathrm\Gamma(\alpha)$ in the complex plane $\mathbb C$ is determined
by the equations
$$
\mathrm\Gamma_\alpha: \quad x=\mathrm{Re}\frac{\Lambda^+(\mu)}
{\Lambda^-(\mu)}, \quad y=\mathrm{Im}\frac{\Lambda^+(\mu)}
{\Lambda^-(\mu)}, \quad 0\leqslant\mu\leqslant+\infty.
$$

\begin{figure}[t]
\begin{center}
\includegraphics[width=0.47\textwidth]{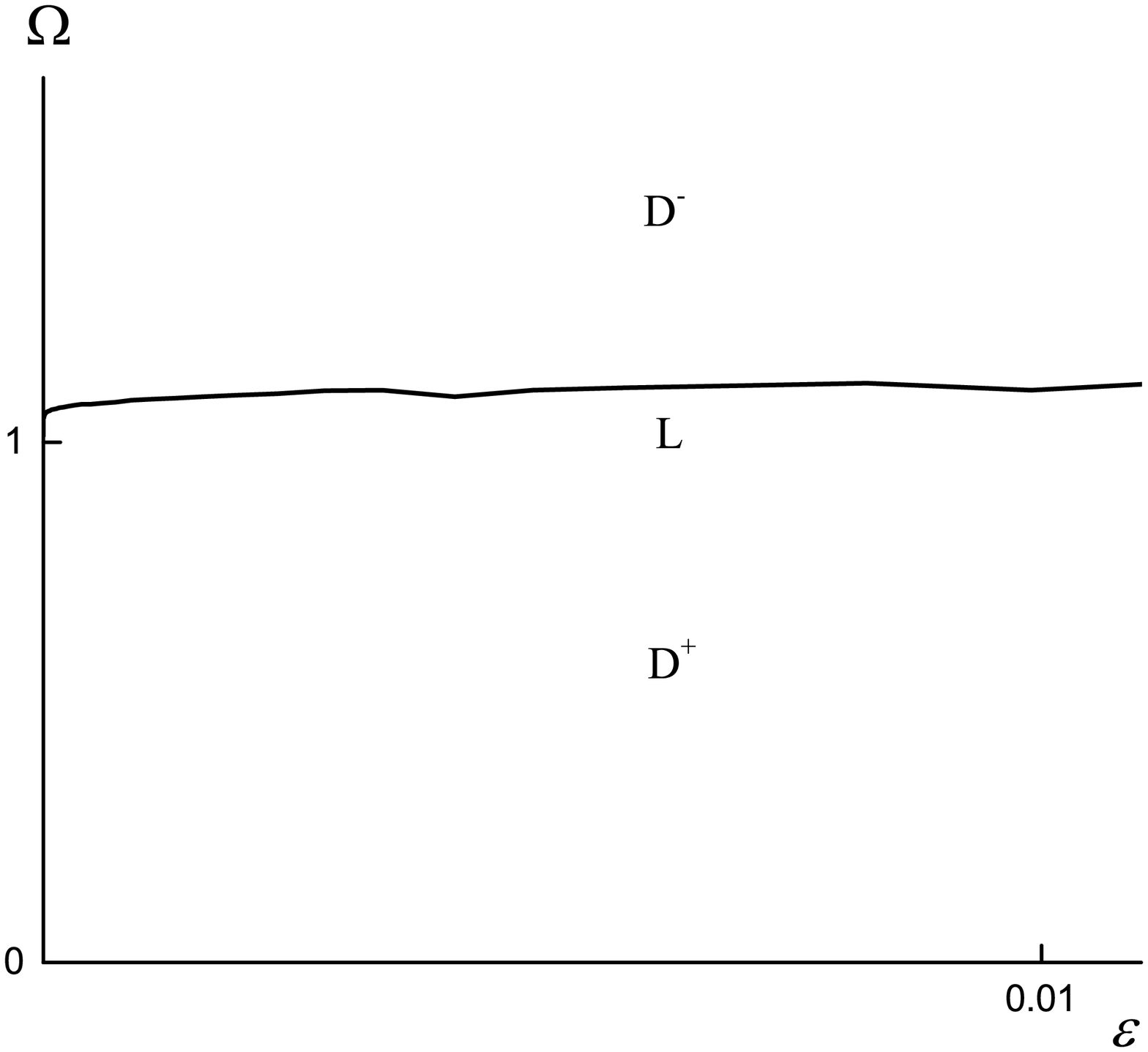}\hfill
\includegraphics[width=0.47\textwidth]{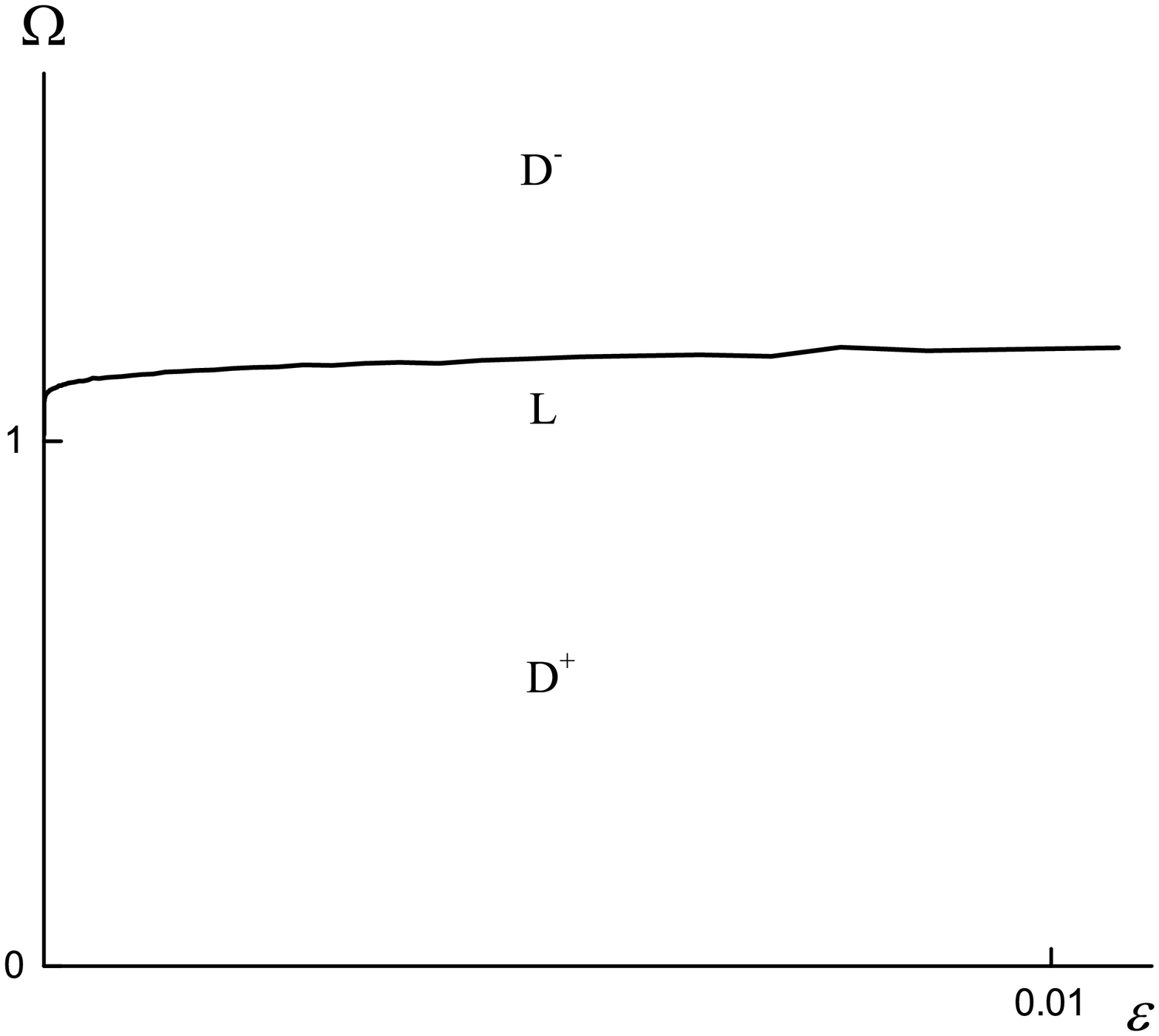}\\
\parbox[t]{0.47\textwidth}{\hspace{3cm}{Fig. 3.}}\hfill
\parbox[t]{0.47\textwidth}{\hspace{3cm}{Fig. 4.}}
\end{center}
\end{figure}

We accentuate that the curve $L_\alpha$ on the plane of the parameters
$(\Omega,\varepsilon)$
defined by the parametric equations
$$
L(\alpha): \quad \mathrm{Re}\frac{\Lambda^+(\mu,\alpha,\Omega,\varepsilon)}
{\Lambda^-(\mu,\alpha,\Omega,\varepsilon)}=0, \quad
\mathrm{Im}\frac{\Lambda^+(\mu,\alpha,\Omega,\varepsilon)}
{\Lambda^-(\mu,\alpha,\Omega,\varepsilon)}=0, \quad 0\leqslant\mu\leqslant+\infty.
$$

The value of the reduced chemical potential $\alpha$ fills the entire real
line $-\infty\leqslant\alpha\leqslant+\infty$.
In this case $\alpha$=$-\infty$ corresponds to a Maxwell plasma and the case
$\alpha$=$+\infty$ corresponds to a completely degenerate plasma.

We formulate conclusions in terms of the plasma (Debye or discrete) mode.
If $(\Omega,\varepsilon)\in D^+(\alpha)$ then plasma mode
$H_{\eta_0}(x_1,\mu), e_{\eta_0}(x_1)$ exists
(the number of zeros of the dispersion function $\Lambda(z)$ is equal to two or
index of coefficient
$G(\mu)=\Lambda^+(\mu)/\Lambda^-(\mu)$ is equal to one in the real semi-axis).
If $(\Omega,\varepsilon)\in D^-(\alpha)$ then plasma mode does not exist
(the number of zeros of the dispersion function is zero or the index of the
coefficient is equal to zero on the real semi-axis).

\begin{center}
{5. SPECULAR REFLECTION OF ELECTRONS FROM THE PLASMA BOUNDARY}
\end{center}

We will solve the problem which consists of equations (1.1) and (1.2) with the
boundary conditions (1.3) - (1.6).
We look for the solution of a problem in the form of expansions
$$
H(x,\mu)=\dfrac{E_\infty}{w_0}\mu+\dfrac{E_0}{w_0}
\dfrac{\eta_0\mu-\eta_1^2}{\eta_0-\mu}
\exp\Big(-\dfrac{w_0x}{\eta_0}\Big)+
\dfrac{1}{w_0}\int_{0}^{\infty}\exp\Big(-\dfrac{w_0x}{\eta}\Big)
F(\eta,\mu)\,E(\eta)\,d\eta,
\eqno{(5.1)}
$$
$$
e(x)=E_\infty+E_0\exp\Big(-\dfrac{w_0x}{\eta_0}\Big)+
\int_{0}^{\infty}\exp\Big(\dfrac{w_0x}{\eta}\Big)E(\eta)\,d\eta.
\eqno{(5.2)}
$$

The unknowns in the expansions (5.1) and (5.2) are the coefficients
discrete spectrum $E_0,\; E_\infty$ and the coefficient of the continuous
spectrum  $E(\eta)$ and
if $(\Omega,\varepsilon)\in D^-(\alpha)$ then $E_0=0$.

Consider the case $(\Omega, \varepsilon) \in D^{+}(\alpha)$. We substitute the
expansions (5.1) and (5.2) into the boundary conditions (1.3) - (1.6).
We obtain the following system of equations

$$
2E_{\infty}\mu + E_0 \left(\frac{\eta_1^2-\eta_0 \mu}{\mu - \eta_0}+
\frac{\eta_1^2+\eta_0 \mu}{\mu + \eta_0}\right) + \int_{0}^{\infty}
\left[F(\eta,\mu)-F(\eta,-\mu)\right] E(\eta)d\eta=0,
\eqno{(5.3)}
$$
$$
E_{\infty}+E_{0}+\int_{0}^{\infty}E(\eta)d\eta=1.
\eqno{(5.4)}
$$

Extending the function $E(\eta)$ into the interval $(-\infty,0)$ evenly,
so that $E(\eta)=E(-\eta)$.
Then we get the following equation $F(-\eta,-\mu)=-F(\eta,\mu)$. We transform
the equation (5.3) to the form
$$
2E_\infty\mu + E_0\left(\frac{\eta_1^2-\eta_0 \mu}{\mu - \eta_0}+
\frac{\eta_1^2+\eta_0 \mu}{\mu + \eta_0}\right) +
\int_{-\infty}^{\infty} F(\eta,\mu)E(\eta)d\eta=0,
\eqno{(5.5)}
$$

We substitute in equation (5.5) the eigenfunctions. We obtain the singular
integral equation with the Cauchy
kernel on the whole real axis $-\infty <\mu <+\infty$
$$
2E_{\infty}\mu + E_0 \left[\frac{\eta_1^2-\eta_0 \mu}{\mu - \eta_0}+
\frac{\eta_1^2+\eta_0 \mu}{\mu + \eta_0}\right] +
\int_{-\infty}^{\infty}\frac{\mu \eta-\eta^2_1}{\eta-\mu}E(\eta)d\eta -
2\eta^2_1 w_0 s_0(\alpha)\frac{\Lambda(\mu,\alpha)}{\mu f_0(\mu,\alpha)}=0.
\eqno{(5.6)}
$$

We introduce the auxilary function

$$
M(z)=\int_{-\infty}^{\infty}\frac{z\eta - \eta^2_1}{\eta - z}E(\eta)d\eta
\eqno{(5.7)}
$$

The function $M(z)$ is analytic in the complex plane without the cut
(the point of integration the whole real axis $(-\infty,+\infty)$). The boundary
values of the function $M(z)$ from above and below by a cut is defined as the limits

$$
M^+(\mu) = \lim_{\substack{z\to \mu, \\ \mathrm{Im} \ z>0}} M(z),
M^-(\mu) = \lim_{\substack{z\to\mu, \\ \mathrm{Im}\ z<0}} M(z),   -\infty < \mu < +\infty.
$$

The boundary values of the auxiliary function $M(z)$ are related by the Sokhotzky
formulas

$$
M^\pm(\mu) =
\pm\pi i(\mu^2-\eta^2_1)E(\mu)+\int\limits_{-\infty}^{\infty}
\frac{\mu\eta-\eta^2_1}{\eta-\mu}E(\eta)d\eta,
$$
where the integral

$$
M(\mu) = \int\limits_{-\infty}^{\infty}\frac{\mu\eta-\eta^2_1}{\eta-\mu}E(\eta) d\eta
$$
is understood as singular in terms of the principal value by Cauchy.

The equations follows from Sokhotzky formulas
$$
M^+(\mu) - M^-(\mu) = 2\pi i (\mu^2 - \eta^2_1)E(\mu), \quad  \mu \in (-\infty, +\infty),
\eqno {(5.8)}
$$

$$
M(\mu) = \frac{M^+(\mu) +  M^-(\mu)}{2}, \quad  \mu \in (-\infty, +\infty).
$$

We transform the singular equation (5.6) with the help of (5.7) and (5.8)
to the boundary value problem of Riemann
$$
\Lambda^+(\mu,\alpha)\left[M^+(\mu)+2E_{\infty}\mu+
E_{0}\left(\frac{\eta_1^2-\eta_0 \mu}{\mu - \eta_0}+
\frac{\eta_1^2+\eta_0 \mu}{\mu + \eta_0}\right)\right]=
$$$$
=\Lambda^-(\mu,\alpha)\left[M^-(\mu)+2E_{\infty}\mu+
E_0\left(\frac{\eta_1^2-\eta_0 \mu}{\mu - \eta_0}+
\frac{\eta_1^2+\eta_0 \mu}{\mu + \eta_0}\right)\right], \quad \mu \in (-\infty,+\infty).
\eqno {(5.9)}
$$

Problem (5.9) has the following solution

$$
M(z)=-2E_{\infty}z-E_0\left(\frac{\eta^2_1-\eta_0z}{z-\eta_0}+
\frac{\eta^2_1+\eta_0z}{z+\eta_0}\right)+\frac{C_1z}{\Lambda(z,\alpha)}.
\eqno {(5.10)}
$$

Let us eliminate the pole of the solution in the infinity. We get that

$$
C_1=2E_{\infty}\Lambda_{\infty}.
$$

The amplitude of Debye is in eliminating poles from solutions (5.10) at the points
$\pm\eta_0$. Since the dispersion function is even then these poles are eliminated
one condition

$$
E_0=\frac{C_1\eta_0}{(\eta^2_1-\eta^2_0)\Lambda'(\eta_0,\alpha)}=
\frac{2E_{\infty}\Lambda_{\infty}\eta_0}{(\eta^2_1-\eta^2_0)\Lambda'(\eta_0,\alpha)}.
$$

If we substitute the solution (5.10) in the Sokhotzky formulas (5.8) then we find
coefficient of the continuous spectrum

$$
E(\mu)=\frac{C_1\mu}{2\pi i(\mu^2-\eta^2_1)}
\left(\frac{1}{\Lambda^+(\mu)-\Lambda^-(\mu)}\right)=\frac{E_{\infty}
\Lambda_{\infty}\varepsilon \mu^2f_0(\mu,\alpha)}{(\varepsilon - i\Omega)
\eta^2_1s_0(\alpha)\Lambda^+(\mu)\Lambda^-(\mu)}.
$$

To find the $E_{\infty}$ we use the equation (5.4), which we rewrite in view
$E(\eta)$ is even:

$$
E_{\infty}+E_0+\frac{1}{2}\int\limits_{-\infty}^{\infty}E(\eta)d\eta=1,
$$
or in the explicit form

$$
\frac{1}{\Lambda_{\infty}}+\frac{2\eta_0}{(\eta^2_1-\eta^2_0)\Lambda'(\eta_0)}+
\frac{1}{2\pi i}\int\limits_{-\infty}^{\infty}
\left(\frac{1}{\Lambda^+(\eta)}-
\frac{1}{\Lambda^-(\eta)}\right)\frac{\eta d\eta}{\eta^2-\eta^2_1}=
\frac{1}{\Lambda_{\infty}E_{\infty}}.
\eqno {(5.11)}
$$

The integral from (5.11)

$$
J=\frac{1}{2\pi i}\int\limits_{-\infty}^{\infty}
\left(\frac{1}{\Lambda^+(\eta)}-
\frac{1}{\Lambda^-(\eta)}\right)\frac{\eta d\eta}{\eta^2-\eta^2_1}.
$$
can be calculated analytically. The function

$$
\varphi(z)=\frac{z}{\lambda(z)(z^2-\eta^2_1)},
$$
for which $\varphi(z)$=$O(z^{-1})$ $(z\to\infty)$ is analytic in the complex
plane without the cut with the exception of points $\pm\eta_1$, $\pm\eta_0$.
Consequently, this integral is equal to

$$
J=\Big[\Res_{\eta_0}+\Res_{-\eta_0}+\Res_{\eta_1}+\Res_{-\eta_1}+
\Res_{\infty}\Big]\varphi(z).
$$
Observing that

$$
\Res_{\pm \eta_1}\varphi(z)=\dfrac{1}{2\Lambda_1},
 \Lambda_1=\Lambda(\eta_1)=1-\frac{\varepsilon}{\varepsilon - i\Omega},
$$

$$
\Res_{\pm \eta_0}\varphi(z)=\frac{\eta_0}{\Lambda'(\eta_0)(\eta^2_0-\eta^2_1)},
$$
we obtain

$$
J=\frac{2\eta_0}{\Lambda'(\eta_0)(\eta^2_0-\eta^2_1)}+
\frac{1}{\Lambda_1}-\frac{1}{\Lambda_{\infty}}.
$$

Substituting this equality into (5.11) we find that

$$
E_{\infty} = \frac{\Lambda_1}{\Lambda_{\infty}},   C_1=2\Lambda_1.
$$

Thus, the expansion (5.1) and (5.2) are found. The structure of the electric
field generally as follows

$$
e(x)=\frac{\Lambda_1}{\Lambda_{\infty}}+
\frac{2\Lambda_1\eta_0\mathrm{exp}(-{w_0}x/\eta_0)}{\Lambda'(\eta_0,\alpha)
(\eta^2_1-\eta^2_0)}
+\frac{\Lambda_1}{{w_0}\eta^2_1 s_0(\alpha)}\int\limits_{0}^{\infty}
\frac{\eta^2 f_0(\eta,\alpha)\mathrm{exp}(-{w_0}x/\eta)}{\Lambda^+(\eta,\alpha)
\Lambda^-(\eta,\alpha)}d\eta.
\eqno {(5.12)}
$$

Recall that the formula (5.12) holds for $(\Omega,\varepsilon)\in D^+(\alpha)$.
In the case $(\Omega,\varepsilon)\in D^-(\alpha)$ zero $\eta_0$ of the dispersion
function does not exist. Therefore, we can assume that in this case $\eta_0=0$.
Then the second term in (5.12) vanishes and the formula (5.12) is simplified.

Here is also an explicit expansion of the distribution function $H(x,\mu)$

$$
H(x,\mu)=\frac{\Lambda_1}{{w_0}\Lambda_{\infty}}\mu +
\frac{2\Lambda_1\eta_0}{{w_0}(\eta^2_1-\eta^2_0)
\Lambda'(\eta_0,\alpha)}\frac{\eta_0\mu-\eta^2_1}{\eta_0-\mu}\mathrm{exp}
\left(-\frac{w_0 x}{\eta_0}\right)+
$$

$$
+\frac{\Lambda_1}{{w_0}^2 \eta^2_1s_0(\alpha)}\int\limits_{0}^{\infty}
\mathrm{exp}\left(-\frac{w_0 x}{\eta_0}\right) \frac{F(\eta,\mu)\eta^2
f_0(\eta,\alpha)d\eta}{\Lambda^+(\eta,\alpha)\Lambda^-(\eta,\alpha)}.
\eqno {(5.13)}
$$

For the first time the problem of oscillations of electron plasma  with a
basic equilibrium Fermi--Dirac statistics in a half-space with specular
boundary conditions formulated and solved analytically.

\end{document}